# The Bessel zeta function

M.G. Naber[a]
mnaber@monroeccc.edu
Department of Science and Mathematics
Monroe County Community College
1555 S. Raisinville Rd
Monroe, Michigan, 48161-9746

B.M. Bruck
brittanie.Bruck@rockets.utoledo.edu
Department of Mathematics and Statistics
University of Toledo
2801 W. Bancroft St.
Toledo, OH, 43606

S.E. Costello
shannon.naki@rockets.utoledo.edu
Department of Physics
University of Toledo
2801 W. Bancroft St.
Toledo, OH, 43606

[a] Author to whom correspondence should be addressed.

**ABSTRACT**

Two representations of the Bessel zeta function are investigated. An incomplete representation is constructed using contour integration and an integral representation due to Hawkins is fully evaluated (analytically continued) to produce two infinite series. This new representation, evaluated at integer values of the argument, produces results that are consistent with known results (values, slope, and pole structure). Not surprisingly, the two representations studied are found to have similar coefficients but a slightly different functional form. A representation of the Riemann zeta function is obtained by allowing the order of the Bessel function to go to 1/2.

**I. INTRODUCTION**

In this brief report properties of the Bessel zeta function are examined. An incomplete representation is found using a contour integration method. A complete representation is found by evaluating an integral representation first obtained by Hawkins[1]. Evaluating the Hawkins integral is computationally simpler than prior investigations (see Refs. 1, 2, 3, and 4) and results in two infinite series of known quantities. The Bessel zeta function is defined as follows: let $J_\nu(x)$ denote the Bessel function of the first kind of order $\nu$, denote by $j_{\nu,n}(n = 1,2,3,...)$ as the zeros of the Bessel function, $J_\nu(j_{\nu,n}) = 0$ (in this paper $\nu > -1$ shall be assumed, in which case the $j_{\nu,n}$ can be taken to be real and positive, the root equal to zero is excluded), and the zeros are assumed to be ordered, $0 < j_{\nu,1} < j_{\nu,2} < j_{\nu,3}$ .... Then,

$$\zeta_\nu(s) = \sum_{n=1}^{\infty} (j_{\nu,n})^{-s}. \qquad (1)$$

For the special case of $\nu = 1/2$, Eq. (1) reduces to a function proportional to the Riemann zeta function

$$\zeta_{1/2}(s) = \pi^{-s}\zeta(s), \qquad (2)$$

where $\zeta(s)$ is the Riemann zeta function[5],

$$\zeta(s) = \sum_{n=1}^{\infty} n^{-s}. \qquad (3)$$

This is because $J_{1/2}(x) = \sqrt{2/\pi x}\sin(x)$.

The physical importance of these studies is broad. This would include essentially any problem where the Laplacian can be expressed, due to symmetry, in cylindrical or spherical coordinates: diffusion, vibration, wave propagation, or the Schrödinger equation (see for example Refs. 3 and 6 and references therein, see also Ref.7 for applications to the Casimir effect).

Section II will go over previous Bessel zeta function results and their computational methods. In section III, a contour integration method will be used to obtain a representation of the Bessel zeta function; however, this will be of limited use as not all the integrals on the contour are able to be evaluated. The results of this section coupled with the final form of the Bessel zeta function from section IV may provide results for integrals containing Bessel functions. In section IV, an integral representation due to Hawkins will be evaluated to provide a new representation of the Bessel zeta function. This representation will give the Bessel zeta function in terms of two infinite sums of known quantities. The coefficients of these series can be related to the coefficients of the series found in section III. In section V this new representation and its derivative will be evaluated at integer values of the argument. It is found that the results agree with the known values of the Bessel zeta function. The slope is also found to be bounded everywhere but at the poles. In the appendix a representation of the Riemann zeta function is obtained by letting the order of the Bessel function go to $1/2$ and then noting a restriction on the domain due to the poles in the Bessel zeta function.

**II. PREVIOUS RESULTS**

There is a long history of investigations of the sums of the even powers of the zeros of Bessel functions going back to Euler[8] (See also G.N. Watson[9] pp. 500-501 for an English language version of the derivations) and Rayleigh[10]. These sums have taken on the name Rayleigh functions[11,12]. In what might be referred to as "pre-zeta function" results, Sneddon[13] used Fourier-Bessel and Dini series coupled with known Bessel function identities to obtain formulas for the sums of even powers of Bessel zeros and the zeros of derivatives of the Bessel functions. Written in terms of the Bessel zeta function (Eq. 1), the following formula can be obtained[3,9,12,13],

$$\zeta_\nu(2) = \frac{1}{2^2(\nu+1)}, \tag{4}$$

$$\zeta_\nu(4) = \frac{1}{2^4(\nu+1)^2(\nu+2)}, \tag{5}$$

$$\zeta_\nu(6) = \frac{1}{2^5(\nu+1)^3(\nu+2)(\nu+3)}, \tag{6}$$

$$\zeta_\nu(8) = \frac{5\nu+11}{2^8(\nu+1)^4(\nu+2)^2(\nu+3)(\nu+4)}, \tag{7}$$

$$\zeta_\nu(10) = \frac{7\nu+19}{2^9(\nu+1)^5(\nu+2)^2(\nu+3)(\nu+4)(\nu+5)}, \tag{8}$$

etc.

and, the so-called linear and quadratic recursion formula,

$$\sum_{k=0}^{n} \frac{(-1)^k 4^k \zeta_\nu(2k+2)}{(n-k)!\,\Gamma(n+\nu+1-k)} = \frac{1}{4n!\,\Gamma(n+2+\nu)}, \quad n = 0,1,\ldots \tag{9}$$

$$\zeta_\nu(2n) = \frac{1}{n+\nu} \sum_{k=1}^{n-1} \zeta_\nu(2k)\zeta_\nu(2n-2k), \quad n = 2,3,\ldots. \tag{10}$$

$\zeta_\nu(0)$ falls outside of the above sequence and is given by (see pg. 26 of Ref. 1)

$$\zeta_\nu(0) = -\frac{1}{2}\left(\nu + \frac{1}{2}\right). \tag{11}$$

Hawkins[1] appears to be the first to systematically study the Bessel zeta function (not just the sums of even powers of the zeros of the Bessel functions or their derivatives). Hawkins examined an integral representation of the Bessel zeta function,

$$\zeta_\nu(s) = \frac{s}{\pi} \sin(s\pi/2) \int_0^\infty \ln\left(\frac{2^\nu \Gamma(\nu+1)}{x^\nu} I_\nu(x)\right) x^{-s-1} dx. \tag{12}$$

He did not evaluate the integral; but rather, used it to prove analytic continuation properties of the zeta function. He also produced another representation by generalizing the greatest integer function and viewing the Bessel zeta function as a perturbation of the Riemann zeta function. Hawkins was able to verify Eqs. (4)-(11) and was able to show that the Bessel zeta function is analytic in the half plane $Re(s) > -1$, and has a meromorphic continuation to the (complex) $s$ plane where the only possible singularities are simple poles at $s = 1, -1, -3, \ldots$ (see also Eq. (3.19) of Ref. 4) with residues,

$$Res(s = 1) = \frac{1}{\pi}, \tag{13}$$

$$Res(s = -2m - 1) = \frac{(-1)^{m-1} c_{2m}}{\pi}, m = 0,1,2,\ldots, \tag{14}$$

where,

$$c_0 = c_1 = \frac{1}{2}\left(\nu^2 - \frac{1}{4}\right), \tag{15}$$

$$c_2 = -\frac{1}{2^2}\left(\nu^2 - \frac{25}{4}\right) c_0, \tag{16}$$

$$c_3 = -\left(\nu^2 - \frac{13}{4}\right) c_0, \tag{17}$$

$$c_4 = \frac{1}{2^3}\left(\nu^4 - \frac{57}{2}\nu^2 + \frac{1073}{16}\right) c_0, \tag{18}$$

$$c_5 = \left(\nu^4 - \frac{53}{4}\nu^2 + \frac{103}{4}\right) c_0, \tag{19}$$

etc.

In general,

$$c_{m+2} = \frac{1}{2}\left((m+3) c_{m+1} - \sum_{k=0}^{m} c_{m-k} c_k\right). \tag{20}$$

Hawkins was also able to use the analytic continuation of the Bessel zeta function to show,

$$\zeta_\nu(-2m) = \frac{(-1)^m c_{2m-1}}{2}, \ m = 1,2,\ldots. \tag{21}$$

Leseduarte and Romeo[4] obtained the same results for $\zeta_\nu(-2m)$ as Hawkins using heat kernel methods and then again using a contour integration method.

Stolarsky[14] was later able to show the following interesting limit involving the $c_m$,

$$\lim_{m\to\infty} \frac{c_m(\nu)}{c_m(\nu=0)} = cos(\nu\pi). \tag{22}$$

This is a result that was speculated at by Hawkins but left unproven (see pg. 53 of Ref. 1).

Actor and Bender expanded on Hawkins' methods and were able to obtain the derivative of the Bessel zeta function at the origin,

$$\zeta'_\nu(0) = \frac{1}{2} ln\left(\frac{2^\nu \Gamma(\nu+1)}{\sqrt{2\pi}}\right). \tag{23}$$

If $\nu = 1/2$ this does reduce to the known Riemann zeta function result,

$$\zeta'(0) = -\frac{1}{2} ln(2\pi). \tag{24}$$

### III. BESSEL ZETA FUNCTION VIA CONTOUR INTEGRATION

Frequently a contour integral can provide an elegant means of obtaining a representation of zeta functions, provided the resulting integrals can be evaluated[15,16]. This method also allows for the zeta function in question to be analytically continued and thus evaluated at points where the sum would otherwise be divergent. Recall the following result from complex analysis. Let $f(z)$ have zeros at $z_1, z_2, z_3, \ldots$. Let $\gamma$ be a simple closed curve that captures all the zeros of $f(z)$ in a counterclockwise sense. Then,

$$\frac{1}{2\pi i}\oint_\gamma z^s \frac{f'(z)}{f(z)} dz = \sum_{n=1}^\infty (z_n)^s. \tag{25}$$

Let $f(z) = J_\nu(z)$ and then $z_n = j_{\nu,n}$ are the real positive roots of Bessel functions (take $\nu > -1$; then the zeros of the Bessel function are guaranteed to be real.). The zero at the origin will need to be excluded so that there is no singularity in the zeta function. The path, $\gamma$, should not cross the negative real axis as this is a branch cut for the function $z^s$. Take the path $\gamma$ to start on the real axis between the origin and the first positive zero of the Bessel function. The path then travels just below the real axis out to infinity and then just above the real axis back to where it started between the origin and the first positive zero of the Bessel function. Changing $s$ to $-s$ gives the Bessel zeta function.

$$\zeta_\nu(s) = \sum_{n=1}^{\infty} (j_{\nu,n})^{-s} = \frac{1}{2\pi i} \oint_\gamma z^{-s} \frac{J_\nu'(z)}{J_\nu(z)} dz. \tag{26}$$

The contour $\gamma$ can be deformed to a new one provided no poles or branch cuts are crossed. The deformed contour, $\gamma'$, has four pieces: a semi-circular arc that starts at $\theta = -\frac{\pi}{2}$ and ends at $\theta = \frac{\pi}{2}$, the radius of this semi-circular arc goes out to infinity (outer arc), there is another semi-circular arc near the origin that starts at $\theta = \frac{\pi}{2}$ and ends at $\theta = -\frac{\pi}{2}$, the radius of this arc is allowed to go to zero (inner arc), another piece of the contour starts at $z = i\infty$ and goes to $i\varepsilon$ along the positive imaginary axis, $\varepsilon$ is allowed to go to zero, the last piece of the contour starts at $-i\varepsilon$ and goes to $-i\infty$, $\varepsilon$ is allowed to go to zero. This contour avoids the Bessel root at the origin and captures all the positive zeros of the Bessel function. Eq. (26) then reduces to,

$$\zeta_\nu(s) = \frac{1}{2\pi i} \oint_{\gamma'} z^{-s} \frac{J_\nu'(z)}{J_\nu(z)} dz = Z_\nu(s) + W_\nu(s), \tag{27}$$

where,

$$Z_\nu(s) = \frac{1}{2\pi i} \int_{i\infty}^{0} z^{-s} \frac{J_\nu'(z)}{J_\nu(z)} dz + \frac{1}{2\pi i} \int_{0}^{-i\infty} z^{-s} \frac{J_\nu'(z)}{J_\nu(z)} dz, \tag{28}$$

and,

$$W_\nu(s) = \frac{1}{2\pi i} \lim_{R \to \infty} \int_{-\frac{\pi}{2}}^{\frac{\pi}{2}} z^{-s} \frac{J_\nu'(z)}{J_\nu(z)} dz + \frac{1}{2\pi i} \lim_{\epsilon \to 0} \int_{\frac{\pi}{2}}^{-\frac{\pi}{2}} z^{-s} \frac{J_\nu'(z)}{J_\nu(z)} dz, \tag{29}$$

and $z = Re^{i\theta}$ in the first integral of Eq. (29) and $z = \epsilon e^{i\theta}$ in the second.

The integrals defining the $Z_\nu(s)$ function can be merged into one using a Bessel identity and coordinate change to become,

$$Z_\nu(s) = \frac{\sin\left(s\frac{\pi}{2}\right)}{2\pi} \int_0^\infty y^{-s} \frac{I_{\nu-1}(y) + I_{\nu+1}(y)}{I_\nu(y)} dy. \tag{30}$$

This integral can be evaluated by breaking it at $y = 1$ and using appropriate power series about the origin and infinity.

$$\int_0^\infty y^{-s} \frac{I_{\nu-1}(y) + I_{\nu+1}(y)}{I_\nu(y)} dy$$
$$= \int_0^1 y^{-s} \frac{I_{\nu-1}(y) + I_{\nu+1}(y)}{I_\nu(y)} dy + \int_1^\infty y^{-s} \frac{I_{\nu-1}(y) + I_{\nu+1}(y)}{I_\nu(y)} dy. \tag{31}$$

A series about infinity (the second integral) can be constructed, using MAPLE,

$$\frac{I_{v-1}(y) + I_{v+1}(y)}{I_v(y)} = 2 - \frac{1}{y} + \left(v^2 - \frac{1}{4}\right)\sum_{n=2}^{\infty}\frac{d_n}{y^n}, \tag{32}$$

where,

$$d_{m+2} = \frac{m+1}{2}d_{m+1} + \frac{1-4v^2}{16}\sum_{k=0}^{m-2}d_{m-k}d_{k+2}. \tag{33}$$

The above formula, Eq. (33), is for $m \geq 2$ and $v \neq 1/2$

$$d_2 = 1, \tag{34}$$

$$d_3 = 1, \tag{35}$$

$$d_4 = -\frac{(4v^2 - 25)}{16}, \tag{36}$$

$$d_5 = -\frac{(4v^2 - 13)}{4}, \tag{37}$$

$$d_6 = \frac{16v^4 - 456v^2 + 1073}{128}, \tag{38}$$

$$d_7 = \frac{4v^4 - 53v^2 + 103}{4}, \tag{39}$$

$$d_8 = -\frac{320v^6 - 24560v^4 + 218812v^2 - 375733}{4096}, \tag{40}$$

$$d_9 = -\frac{64v^6 - 2160v^4 + 15084v^2 - 23797}{64}, \tag{41}$$

etc.

When $v = 1/2$ only the first two terms are left in the above series (Eq. (32)). With the $v^2 - 1/4$ part factored out the series the Bessel component is separated from the Riemann component of this piece of the zeta function. The second integral on the righthand side of Eq. (31) can now be evaluated

$$\int_1^{\infty} y^{-s}\frac{I_{v-1}(y) + I_{v+1}(y)}{I_v(y)}dy = \int_1^{\infty}\left(2y^{-s} - y^{-s-1} + \left(v^2 - \frac{1}{4}\right)\sum_{n=2}^{\infty}d_n y^{-s-n}\right)dy. \tag{42}$$

The first term on the right-hand side necessitates that $s > 1$. Then,

$$\int_1^\infty y^{-s} \frac{I_{\nu-1}(y) + I_{\nu+1}(y)}{I_\nu(y)} dy = \frac{2}{s-1} - \frac{1}{s} + \left(\nu^2 - \frac{1}{4}\right) \sum_{n=2}^\infty \frac{d_n}{s+n-1}. \quad (43)$$

For the first integral on the righthand side of Eq. (31) a series about the origin is needed. Recall,

$$I_\nu(y) = \left(\frac{y}{2}\right)^\nu \sum_{n=0}^\infty \frac{1}{n!\,\Gamma(n+\nu+1)} \left(\frac{y}{2}\right)^{2n}. \quad (44)$$

Then,

$$\frac{I_{\nu-1}(y) + I_{\nu+1}(y)}{I_\nu(y)} = \frac{\sum_{n=0}^\infty \frac{1}{n!\,\Gamma(n+\nu)} \left(\frac{y}{2}\right)^{2n-1} + \sum_{n=0}^\infty \frac{1}{n!\,\Gamma(n+\nu+2)} \left(\frac{y}{2}\right)^{2n+1}}{\sum_{n=0}^\infty \frac{1}{n!\,\Gamma(n+\nu+1)} \left(\frac{y}{2}\right)^{2n}}. \quad (45)$$

Note that the numerator is an odd function while the denominator is an even function. Hence, the resulting series must be odd. Combining the two series in the numerator gives,

$$= \frac{\frac{1}{\Gamma(\nu)} \left(\frac{y}{2}\right)^{-1} + \sum_{n=0}^\infty \left(\frac{(2n+\nu+2)}{(n+1)!\,\Gamma(n+\nu+2)}\right) \left(\frac{y}{2}\right)^{2n+1}}{\sum_{n=0}^\infty \frac{1}{n!\,\Gamma(n+\nu+1)} \left(\frac{y}{2}\right)^{2n}}. \quad (46)$$

To be able to evaluate the integral, Eq. (46) needs to be expressed as a single series,

$$\frac{I_{\nu-1}(y) + I_{\nu+1}(y)}{I_\nu(y)} = \sum_{n=0}^\infty a_n \left(\frac{y}{2}\right)^{2n-1}. \quad (47)$$

To solve for the $a_n$, denote the two known series as follows,

$$\frac{\frac{1}{\Gamma(\nu)} \left(\frac{y}{2}\right)^{-1} + \sum_{n=0}^\infty b_n \left(\frac{y}{2}\right)^{2n+1}}{\sum_{n=0}^\infty c_n \left(\frac{y}{2}\right)^{2n}}, \quad (48)$$

where

$$b_n = \frac{(2n+\nu+2)}{(n+1)!\,\Gamma(n+\nu+2)}, \quad (49)$$

and

$$c_n = \frac{1}{n!\,\Gamma(n+\nu+1)}. \qquad (50)$$

Please note the $c_n$ in Eq. (50) are different from the $c_n$ of Sec. II. In this notation Eq. (47) becomes,

$$\frac{\frac{1}{\Gamma(\nu)}\left(\frac{y}{2}\right)^{-1} + \sum_{n=0}^{\infty} b_n \left(\frac{y}{2}\right)^{2n+1}}{\sum_{n=0}^{\infty} c_n \left(\frac{y}{2}\right)^{2n}} = \sum_{n=0}^{\infty} a_n \left(\frac{y}{2}\right)^{2n-1}, \qquad (51)$$

or,

$$\frac{1}{\Gamma(\nu)}\left(\frac{y}{2}\right)^{-1} + \sum_{n=0}^{\infty} b_n \left(\frac{y}{2}\right)^{2n+1} = \left(\sum_{n=0}^{\infty} a_n \left(\frac{y}{2}\right)^{2n-1}\right)\left(\sum_{n=0}^{\infty} c_n \left(\frac{y}{2}\right)^{2n}\right). \qquad (52)$$

Expanding the righthand side of Eq. (52), equating like powers of $y/2$, and solving sequentially gives,

$$a_0 = \nu, \qquad (53)$$

$$a_1 = \frac{2}{(\nu+1)}, \qquad (54)$$

$$a_2 = \frac{-2}{(\nu+2)(\nu+1)^2}, \qquad (55)$$

$$a_3 = \frac{4}{(\nu+3)(\nu+2)(\nu+1)^3}, \qquad (56)$$

$$a_4 = -\frac{10\nu + 22}{(\nu+4)(\nu+3)(\nu+2)^2(\nu+1)^4}, \qquad (57)$$

etc.

A general formula (for $n \geq 1$) is,

$$a_n = \left(b_{n-1} - \sum_{k=0}^{n-1} a_k c_{n-k}\right)\Gamma(\nu+1), \qquad (58)$$

or

$$a_n = \Gamma(\nu+1)\left(\frac{(2n+\nu)}{n!\,\Gamma(n+\nu+1)} - \sum_{k=0}^{n-1} \frac{a_k}{(n-k)!\,\Gamma(n+\nu+1-k)}\right). \qquad (59)$$

Notice that the $a_n$ can be related to the sums of even powers of the Bessel zeros (compare Eqs. (53)-(59) with Eqs. (4)-(10)),

$$a_n = (-1)^{n+1} 2^{2n+1} \zeta_\nu(2n). \tag{60}$$

Note that for the special case of $\nu = 1/2$, the $a_n$ can be written in terms of the Bernoulli numbers (see pg. 2 of Ref. 17 for a definition of Bernoulli numbers) by the following formula:

$$a_n = \frac{2^{4n}}{(2n)!} B_{2n}. \tag{61}$$

The first integral on the righthand side of Eq. (31) can now be evaluated.

$$\int_0^1 y^{-s} \frac{I_{\nu-1}(y) + I_{\nu+1}(y)}{I_\nu(y)} dy = \int_0^1 y^{-s} \sum_{n=0}^\infty \frac{a_n}{2^{2n-1}} y^{2n-1} dy. \tag{62}$$

The first term of the power series necessitates that $s > 0$. Then (interchanging the sum and the integral),

$$\int_0^1 y^{-s} \frac{I_{\nu-1}(y) + I_{\nu+1}(y)}{I_\nu(y)} dy = \sum_{n=0}^\infty \frac{a_n}{2^{2n-1}(2n-s)}, \tag{63}$$

The function $Z_\nu(s)$ can now be expressed as:

$$Z_\nu(s) = \frac{\sin\left(s\frac{\pi}{2}\right)}{2\pi} \left( \sum_{n=1}^\infty \frac{a_n}{2^{2n-1}(2n-s)} + \frac{2}{s-1} - \frac{1+2\nu}{s} + \left(\nu^2 - \frac{1}{4}\right) \sum_{n=2}^\infty \frac{d_n}{s+n-1} \right). \tag{64}$$

This function, $Z_\nu(s)$, by itself, agrees with known results for the Bessel zeta function at $s = 0$, $\pm 2k$ for $k = 1,2, ...$, and has the same residue at $s = 1, -1, -3, ...$ as the Bessel zeta function; however, it does not have the same slope at the origin as the Bessel zeta function is known to have[2],

$$Z'_\nu(0) \neq \zeta'_\nu(0) = \frac{1}{2} \ln\left(\frac{2^\nu \Gamma(\nu+1)}{\sqrt{2\pi}}\right). \tag{65}$$

This means there is some key information still to be obtained in the integrals defining $W_\nu(s)$. Unfortunately, those integrals are difficult to evaluate due to the limit process and the form of the integrands. With the two arcs parameterized in the usual polar coordinates, $W_\nu$ takes the form of,

$$2\pi W_\nu(s) = \lim_{R \to \infty} \int_{-\frac{\pi}{2}}^{\frac{\pi}{2}} \frac{R^{1-s}}{e^{i(s-1)\theta}} \frac{J_\nu'(Re^{i\theta})}{J_\nu(Re^{i\theta})} d\theta - \lim_{\epsilon \to 0} \int_{-\frac{\pi}{2}}^{\frac{\pi}{2}} \frac{\epsilon^{1-s}}{e^{i(s-1)\theta}} \frac{J_\nu'(\epsilon e^{i\theta})}{J_\nu(\epsilon e^{i\theta})} d\theta. \tag{66}$$

Recall that since $v > -1$, the zeros of the Bessel function are real, so the denominators of the integrands are never zero. Hence, for a fixed $R$ and $\epsilon$, the integrals are clearly bounded. Recall that Hawkins was able to show that the only possible locations for poles for the Bessel zeta function are at $s = 1, -1, -3, \ldots$. The known residues for the Bessel zeta function are the same as the residues for the $Z_v(s)$ function. This means that the $W_v(s)$ function has no poles. Additionally, the values of the Bessel zeta function and the $Z_v(s)$ are the same at $s = 0$, and $\pm 2k$ for $k = 1,2,3, \ldots$. This means that $W_v(0) = W_v(\pm 2k) = 0$. This information indicates that $W_v(s)$ can be represented as,

$$W_v(s) = \frac{1}{2\pi} \sin\left(s\frac{\pi}{2}\right) G_v(s), \tag{67}$$

where $G_v(s)$ has no poles but may have zeros. The factor of $2\pi$ is there for convenience. The zeta function is then,

$$\zeta_v(s) = \frac{\sin\left(s\frac{\pi}{2}\right)}{2\pi} \left( \sum_{n=1}^{\infty} \frac{2^{1-2n} a_n}{(2n-s)} + \frac{2}{s-1} - \frac{1+2v}{s} + \left(v^2 - \frac{1}{4}\right) \sum_{n=2}^{\infty} \frac{d_n}{s+n-1} + G_v \right). \tag{68}$$

Leseduarte and Romeo[4] also used a contour integral to investigate the Bessel zeta function. Their contour was different than the one used above. They were also able to verify the known pole structure and values of the Bessel zeta function at the even integers. Their final result, however, is not easily evaluated for non-integer values of $s$.

## IV. EVALUATION OF THE HAWKINS REPRESENTATION

Following Hawkins[1], an integral representation of the Bessel zeta function can be obtained from an identity for the modified Bessel function, $I_v$, and an integral representation of the cosecant function. The identities being,

$$\prod_{n=1}^{\infty} \left(1 + \left(\frac{x}{j_{v,n}}\right)^2\right) = \frac{2^v \Gamma(v+1)}{x^v} I_v(x), \tag{69}$$

and,

$$\int_0^{\infty} \ln(1+x^2) x^{-s-1} dx = \frac{\pi}{s} \csc\left(\frac{s\pi}{2}\right). \tag{70}$$

Starting with Eq. (70), scale $x$ by a Bessel zero, $x \to x/j_{v,n}$, and rearranging gives,

$$(j_{v,n})^{-s} = \frac{s}{\pi} \sin(s\pi/2) \int_0^{\infty} \ln\left(1 + \left(\frac{x}{j_{v,n}}\right)^2\right) x^{-s-1} dx. \tag{71}$$

Now sum over all non-zero Bessel roots to obtain,

$$\zeta_\nu(s) = \frac{s}{\pi}\sin\left(\frac{s\pi}{2}\right)\sum_{n=1}^{\infty}\int_0^{\infty}\ln\left(1+\left(\frac{x}{j_{\nu,n}}\right)^2\right)x^{-s-1}dx, \tag{72}$$

or,

$$\zeta_\nu(s) = \frac{s}{\pi}\sin\left(\frac{s\pi}{2}\right)\int_0^{\infty}\ln\left(\prod_{n=1}^{\infty}\left(1+\left(\frac{x}{j_{\nu,n}}\right)^2\right)\right)x^{-s-1}dx. \tag{73}$$

Now substitute Eq. (69) in the integrand to obtain an integral representation of the Bessel zeta function,

$$\zeta_\nu(s) = \frac{s}{\pi}\sin\left(\frac{s\pi}{2}\right)\int_0^{\infty}\ln\left(\frac{2^\nu\Gamma(\nu+1)}{x^\nu}I_\nu(x)\right)x^{-s-1}dx. \tag{74}$$

At this point Hawkins integrates by parts and uses the remaining integral to obtain analytic continuation properties. The above integral can, however, be evaluated. First break the integral at $x = 1$,

$$\int_0^{\infty}\ln\left(\frac{2^\nu\Gamma(\nu+1)}{x^\nu}I_\nu(x)\right)x^{-s-1}dx = \int_0^{1}\ln\left(\frac{2^\nu\Gamma(\nu+1)}{x^\nu}I_\nu(x)\right)x^{-s-1}dx$$
$$+ \int_1^{\infty}\ln\left(\frac{2^\nu\Gamma(\nu+1)}{x^\nu}I_\nu(x)\right)x^{-s-1}dx. \tag{75}$$

The first integral on the right-hand side of Eq. (75) can be evaluated with a series about the origin and the second with a series about infinity. The series about the origin for the logarithm is even in $x$ and is given by,

$$\ln\left(\frac{2^\nu\Gamma(\nu+1)}{x^\nu}I_\nu(x)\right) = \sum_{n=1}^{\infty}(-1)^{n+1}\alpha_n x^{2n}, \tag{76}$$

where,

$$\alpha_1 = \frac{1}{2^2(\nu+1)}, \tag{77}$$

$$\alpha_2 = \frac{1}{2^5(\nu+1)^2(\nu+2)}, \tag{78}$$

$$\alpha_3 = \frac{1}{3\cdot 2^5(\nu+1)^3(\nu+2)(\nu+3)}, \tag{79}$$

$$\alpha_4 = \frac{5\nu + 11}{2^{10}(\nu + 1)^4(\nu + 2)^2(\nu + 3)(\nu + 4)}, \tag{80}$$

$$\alpha_5 = \frac{7\nu + 19}{5 \cdot 2^9(\nu + 1)^5(\nu + 2)^2(\nu + 3)(\nu + 4)(\nu + 5)}, \tag{81}$$

etc.

Notice that,

$$\alpha_k = \frac{(-1)^{k+1}}{k \cdot 2^{2k+1}} a_k = \frac{\zeta_\nu(2k)}{k} \quad \text{for } k \geq 1. \tag{82}$$

The first integral on the righthand side of Eq. (75) can now be evaluated. The first term of the series necessitates that $s < 2$. Then (interchanging the sum and the integral),

$$\int_0^1 \ln\left(\frac{2^\nu \Gamma(\nu + 1)}{x^\nu} I_\nu(x)\right) x^{-s-1} dx = \int_0^1 \sum_{n=1}^\infty (-1)^{n+1} \alpha_n x^{2n} \, x^{-s-1} dx, \tag{83}$$

$$\int_0^1 \ln\left(\frac{2^\nu \Gamma(\nu + 1)}{x^\nu} I_\nu(x)\right) x^{-s-1} dx = \sum_{n=1}^\infty \frac{(-1)^{n+1} \alpha_n}{2n - s}. \tag{84}$$

Now consider the second integral on the righthand side of Eq. (75). Using Maple, a series for the logarithm about infinity is found to be,

$$\ln\left(\frac{2^\nu \Gamma(\nu + 1)}{x^\nu} I_\nu(x)\right) = x + \beta_0 + \sum_{n=1}^\infty \beta_n x^{-n} - \left(\nu + \frac{1}{2}\right) \ln(x), \tag{85}$$

where,

$$\beta_0 = \ln\left(\frac{\Gamma(\nu + 1) 2^\nu}{\sqrt{2\pi}}\right). \tag{86}$$

The odd indexed $\beta$'s fall into a pattern,

$$\beta_1 = -\frac{1}{2}\left(\nu^2 - \frac{1}{4}\right), \tag{87}$$

$$\beta_3 = \frac{1}{3 \cdot 2^3}\left(\nu^2 - \frac{25}{4}\right)\left(\nu^2 - \frac{1}{4}\right), \tag{88}$$

$$\beta_5 = -\frac{1}{5 \cdot 2^4}\left(\nu^4 - \frac{57}{2}\nu^2 + \frac{1073}{16}\right)\left(\nu^2 - \frac{1}{4}\right), \tag{89}$$

etc.

and are expressible in terms of the $c_n$'s from Eqs. (15)-(20),

$$\beta_{2k-1} = -\frac{c_{2k-2}}{2k-1}. \tag{90}$$

A similar pattern occurs for the even indexed $\beta$'s,

$$\beta_2 = -\frac{1}{4}\left(v^2 - \frac{1}{4}\right), \tag{91}$$

$$\beta_4 = \frac{1}{8}\left(v^2 - \frac{13}{4}\right)\left(v^2 - \frac{1}{4}\right), \tag{92}$$

$$\beta_6 = -\frac{1}{6}\left(v^4 - \frac{53}{4}v^2 + \frac{103}{4}\right)\frac{1}{2}\left(v^2 - \frac{1}{4}\right), \tag{93}$$

etc.

or,

$$\beta_{2k} = -\frac{c_{2k-1}}{2k}. \tag{94}$$

The second integral on the righthand side of Eq. (75) now takes the form of,

$$= \int_1^\infty \left( x^{-s} + \beta_0 x^{-s-1} + \sum_{n=1}^\infty \beta_n x^{-n-s-1} - \left(v + \frac{1}{2}\right) \ln(x) x^{-s-1} \right) dx. \tag{95}$$

The first term in the integrand necessitates that $s > 1$. Then,

$$\int_1^\infty \ln\left(\frac{2^v \Gamma(v+1)}{x^v} I_v(x)\right) x^{-s-1} dx = \frac{1}{s-1} + \frac{\beta_0}{s} + \sum_{n=1}^\infty \frac{\beta_n}{n+s} - \left(v + \frac{1}{2}\right)\frac{1}{s^2}. \tag{96}$$

Plugging Eq. (84) and Eq. (96) back into Eq. (74) gives,

$$\zeta_v(s) = \frac{\sin\left(\frac{s\pi}{2}\right)}{\pi}\left(\sum_{n=1}^\infty \frac{(-1)^{n+1}\alpha_n s}{2n-s} + \frac{s}{s-1} + \beta_0 + \sum_{n=1}^\infty \frac{\beta_n s}{n+s} - \left(v + \frac{1}{2}\right)\frac{1}{s}\right). \tag{97}$$

Eq. (97) is a representation of the Bessel zeta function. The $\alpha_n$ are related to the coefficients of the power series, about the origin, of $(I_{v-1}(x) + I_{v+1}(x))/I_v(x)$ and the $\beta_n$ to the power series about infinity. This is similar to the phenomena observed with the Airy zeta function[18].

## V. EVALUATION AT KNOWN POINTS

It now remains to verify this representation, Eq. (97), at known points of the Bessel zeta function. First check at the origin. The last term of Eq. (97) is the only one to survive the limit as $s \to 0$,

$$\zeta_\nu(0) = \lim_{s \to 0} \frac{\sin\left(\frac{s\pi}{2}\right)}{\pi}\left(-\left(\nu + \frac{1}{2}\right)\frac{1}{s}\right) = -\frac{1}{2}\left(\nu + \frac{1}{2}\right). \tag{98}$$

This agrees with the known result. Now check at the positive even integers. In this case, the only term to survive the limit as $s \to 2k, k = 1,2, ...,$ is in the first series (the $\alpha$ series),

$$\zeta_\nu(2k) = \lim_{s \to 2k} \frac{\sin\left(\frac{s\pi}{2}\right)}{\pi}\left(\sum_{n=1}^{\infty} \frac{(-1)^{n+1}\alpha_n s}{2n - s}\right). \tag{99}$$

To evaluate this limit, let $s = t + 2k$. Then,

$$\zeta_\nu(2k) = \lim_{t \to 0} \frac{\sin\left(\frac{t\pi}{2} + k\pi\right)}{\pi}\left(\sum_{n=1}^{\infty} \frac{(-1)^{n+1}\alpha_n(t + 2k)}{2n - 2k - t}\right), \tag{100}$$

$$\zeta_\nu(2k) = \lim_{t \to 0} \frac{\sin\left(\frac{t\pi}{2}\right)(-1)^k}{\pi} \sum_{n=1}^{\infty} \frac{(-1)^{n+1}\alpha_n(t + 2k)}{2n - 2k - t}, \tag{101}$$

$$\zeta_\nu(2k) = \lim_{t \to 0} \frac{t(-1)^k}{2}\frac{(-1)^{k+1}\alpha_k 2k}{-t} = k\alpha_k. \tag{102}$$

This agrees with the known result. Now check at the negative even integers. In this case, the only term to survive the limit as $s \to -2k, k = 1,2, ...,$ is in the second series (the $\beta$ series)

$$\zeta_\nu(-2k) = \lim_{s \to -2k} \frac{\sin\left(\frac{s\pi}{2}\right)}{\pi}\sum_{n=1}^{\infty} \frac{\beta_n s}{n + s}, \tag{103}$$

To evaluate this limit let $s = t - 2k$

$$\zeta_\nu(-2k) = \lim_{t \to 0} \frac{\sin\left(t\frac{\pi}{2} - k\pi\right)}{\pi}\sum_{n=1}^{\infty} \frac{\beta_n(t - 2k)}{n + (t - 2k)}, \tag{104}$$

$$\zeta_\nu(-2k) = \lim_{t \to 0} \frac{\sin\left(\frac{t\pi}{2}\right)(-1)^k}{\pi}\sum_{n=1}^{\infty} \frac{\beta_n(t - 2k)}{n - 2k + t}, \tag{105}$$

$$\zeta_\nu(-2k) = \lim_{t \to 0} \frac{t(-1)^k \beta_{2k}(-2k)}{2} \frac{}{t} = (-1)^{k+1} k \beta_{2k}. \tag{106}$$

This agrees with the known result. The next thing to check is the pole structure. There will always be a pole at $s = 1$ with residue of $1/\pi$. The other poles can only occur at negative odd integers, and they are from the second series (the $\beta$ series).

$$Res(s = 1 - 2k) = \lim_{s \to 1-2k} (s - 1 + 2k) \frac{\sin\left(\frac{s\pi}{2}\right)}{\pi} \frac{\beta_{2k-1} s}{2k - 1 + s}, \tag{107}$$

$$Res(s = 1 - 2k) = \frac{(-1)^k}{\pi} \beta_{2k-1}(1 - 2k) = \frac{(-1)^k}{\pi} c_{2k-2}. \tag{108}$$

This agrees with known results. The last known value to check is the derivative at the origin. The slope of the Bessel zeta function is,

$$\zeta'_\nu(s) = \frac{\cos\left(\frac{s\pi}{2}\right)}{2} \left( \sum_{n=1}^\infty \frac{(-1)^{n+1} \alpha_n s}{2n - s} + \frac{s}{s-1} + \ln\left(\frac{\Gamma(\nu+1) 2^\nu}{\sqrt{2\pi}}\right) + \sum_{n=1}^\infty \frac{\beta_n s}{n+s} - \left(\nu + \frac{1}{2}\right)\frac{1}{s} \right)$$
$$+ \frac{\sin\left(\frac{s\pi}{2}\right)}{\pi} \left( \sum_{n=1}^\infty \frac{(-1)^{n+1} 2n \alpha_n}{(2n-s)^2} - \frac{1}{(s-1)^2} + \sum_{n=1}^\infty \frac{n \beta_n}{(n+s)^2} + \left(\nu + \frac{1}{2}\right)\frac{1}{s^2} \right). \tag{109}$$

Now evaluate this at the origin. Dropping terms that clearly go to zero gives,

$$\zeta'_\nu(0) = \lim_{s \to 0} \left( \frac{1}{2}\left( \ln\left(\frac{\Gamma(\nu+1) 2^\nu}{\sqrt{2\pi}}\right) - \left(\nu + \frac{1}{2}\right)\frac{1}{s} \right) + \frac{\sin\left(\frac{s\pi}{2}\right)}{\pi} \left( \left(\nu + \frac{1}{2}\right)\frac{1}{s^2} \right) \right), \tag{110}$$

$$\zeta'_\nu(0) = \frac{1}{2} \ln\left(\frac{\Gamma(\nu+1) 2^\nu}{\sqrt{2\pi}}\right) + \lim_{s \to 0} \left( -\frac{1}{2}\left(\nu + \frac{1}{2}\right)\frac{1}{s} + \frac{1}{2}\left(\nu + \frac{1}{2}\right)\frac{1}{s} \right), \tag{111}$$

$$\zeta'_\nu(0) = \frac{1}{2} \ln\left(\frac{\Gamma(\nu+1) 2^\nu}{\sqrt{2\pi}}\right). \tag{112}$$

The above slope agrees with the known result[2]. Eq. (112) allows for the computation of the product of all the Bessel roots (see pg. 9 of Ref. 17),

$$\prod_{n=1}^\infty j_{\nu,n} = e^{-\zeta'_\nu(0)} = \sqrt{\frac{\sqrt{2\pi}}{\Gamma(\nu+1) 2^\nu}}. \tag{113}$$

Looking at the structure of Eq. (109) there might be concern that the slope becomes unbounded at $s = \pm 2k$ for $k = 1, 2, \ldots$. First examine the slope at $s = 2k$, dropping terms that will clearly be zero and substituting in $s = 2k$ where no issues will occur gives,

$$\zeta'_\nu(2k) = \frac{(-1)^k}{2}\left(\frac{2k}{2k-1} + \ln\left(\frac{\Gamma(\nu+1)2^\nu}{\sqrt{2\pi}}\right) + \sum_{n=1}^{\infty}\frac{\beta_n 2k}{n+2k} - \left(\nu+\frac{1}{2}\right)\frac{1}{2k}\right)$$
$$+ \lim_{s \to 2k}\left(\frac{(-1)^k}{2}\left(\sum_{n=1}^{\infty}\frac{(-1)^{n+1}\alpha_n s}{2n-s}\right) + \frac{\sin\left(\frac{s\pi}{2}\right)}{\pi}\left(\sum_{n=1}^{\infty}\frac{(-1)^{n+1}2n\alpha_n}{(2n-s)^2}\right)\right). \quad (114)$$

Now break the sums inside the limit at $n = k$,

$$\zeta'_\nu(2k) = \frac{(-1)^k}{2}\left(\frac{2k}{2k-1} + \ln\left(\frac{\Gamma(\nu+1)2^\nu}{\sqrt{2\pi}}\right) + \sum_{n=1}^{\infty}\frac{\beta_n 2k}{n+2k} - \frac{\nu+\frac{1}{2}}{2k}\right)$$
$$- \frac{(-1)^k}{2}\left(\sum_{\substack{n=1\\n\neq k}}^{\infty}\frac{(-1)^n\alpha_n k}{n-k}\right) + \lim_{s \to 2k}\left(-\frac{\alpha_k k}{2k-s} - \frac{\sin\left(\frac{s\pi}{2}\right)}{\pi}\frac{(-1)^k 2k\alpha_k}{(2k-s)^2}\right). \quad (115)$$

Looking only at the limit in Eq. (115), let $s = t + 2k$,

$$\lim_{t \to 0}\left(-\frac{\alpha_k k}{2k-t-2k} - \frac{\sin\left(t\frac{\pi}{2}+k\pi\right)}{\pi}\frac{(-1)^k 2k\alpha_k}{(2k-t-2k)^2}\right), \quad (116)$$

$$\lim_{t \to 0}\left(\frac{\alpha_k k}{t} + \frac{\sin\left(t\frac{\pi}{2}\right)(-1)^k}{\pi}\frac{(-1)^{k+1}2k\alpha_k}{(2k-t-2k)^2}\right), \quad (117)$$

$$\lim_{t \to 0}\left(\frac{\alpha_k k}{t} - \frac{k\alpha_k}{t}\right) = 0. \quad (118)$$

The derivative at $s = 2k$ is then,

$$\zeta'_\nu(2k) = \frac{(-1)^k}{2}\left(\frac{2k}{2k-1} + \ln\left(\frac{\Gamma(\nu+1)2^\nu}{\sqrt{2\pi}}\right) - \frac{\nu+\frac{1}{2}}{2k} + \sum_{n=1}^{\infty}\frac{\beta_n 2k}{n+2k} - \sum_{\substack{n=1\\n\neq k}}^{\infty}\frac{(-1)^n\alpha_n k}{n-k}\right). \quad (119)$$

The above sum is clearly bounded. Now examine the slope, Eq. (109), at $s = -2k$. Deleting terms that clearly go to zero and pulling out terms where the limit is clear gives,

$$\zeta'_\nu(-2k) = \frac{(-1)^k}{2}\left(\sum_{n=1}^{\infty}\frac{(-1)^n\alpha_n k}{n+k} + \frac{2k}{2k+1} + \ln\left(\frac{\Gamma(\nu+1)2^\nu}{\sqrt{2\pi}}\right) + \frac{\nu+\frac{1}{2}}{2k}\right)$$
$$+ \lim_{s\to -2k}\left(\frac{(-1)^k}{2}\sum_{n=1}^{\infty}\frac{\beta_n s}{n+s} + \frac{\sin\left(\frac{s\pi}{2}\right)}{\pi}\sum_{n=1}^{\infty}\frac{n\beta_n}{(n+s)^2}\right). \tag{120}$$

Now break the sums inside the limit at $n = 2k$.

$$\zeta'_\nu(-2k) = \frac{(-1)^k}{2}\left(\sum_{n=1}^{\infty}\frac{(-1)^n\alpha_n k}{n+k} + \frac{2k}{2k+1} + \ln\left(\frac{\Gamma(\nu+1)2^\nu}{\sqrt{2\pi}}\right) + \frac{\nu+\frac{1}{2}}{2k} - \sum_{\substack{n=1\\n\neq 2k}}^{\infty}\frac{2k\beta_n}{n-2k}\right)$$
$$+ \lim_{s\to -2k}\left(-\frac{(-1)^k}{2}\frac{\beta_{2k}2k}{2k+s} + \frac{\sin\left(\frac{s\pi}{2}\right)}{\pi}\frac{2k\beta_{2k}}{(2k+s)^2}\right). \tag{121}$$

Looking only at the limit in Eq. (120), let $s = t - 2k$,

$$\lim_{t\to 0}\left(-(-1)^k\frac{\beta_{2k}k}{2k+t-2k} + \frac{\sin\left(t\frac{\pi}{2}-k\pi\right)}{\pi}\frac{2k\beta_{2k}}{(2k+t-2k)^2}\right), \tag{122}$$

$$\lim_{t\to 0}\left(-(-1)^k\frac{\beta_{2k}k}{t} + \frac{\sin\left(t\frac{\pi}{2}\right)(-1)^k}{\pi}\frac{2k\beta_{2k}}{t^2}\right), \tag{123}$$

$$\lim_{t\to 0}\left(-(-1)^k\frac{\beta_{2k}k}{t} + (-1)^k\frac{k\beta_{2k}}{t}\right) = 0. \tag{124}$$

The derivative at $s = -2k$ is then,

$$\zeta'_\nu(-2k) = \frac{(-1)^k}{2}\left(\sum_{n=1}^{\infty}\frac{(-1)^n\alpha_n k}{n+k} + \frac{2k}{2k+1} + \ln\left(\frac{\Gamma(\nu+1)2^\nu}{\sqrt{2\pi}}\right) + \frac{\nu+\frac{1}{2}}{2k}\right.$$
$$\left. - \sum_{\substack{n=1\\n\neq 2k}}^{\infty}\frac{2k\beta_n}{n-2k}\right). \tag{125}$$

Eq. (125) is clearly finite. The only places where the slope becomes undefined is at the poles, as expected.

## VI. CONCLUSSION

In this paper two representations of the Bessel zeta function were examined. Most significantly, an integral representation due to Hawkins was fully evaluated. This new representation, Eq. (97), was found to agree with all known Bessel zeta function results. This representation is in terms of two infinite series whose coefficients are completely known. This form should allow for greater study of the properties of the Bessel zeta function, in particular the non-trivial zeros and computation of zeros for Bessel functions. In the appendix the special case of $\nu = 1/2$ is considered to obtain a representation of the Riemann zeta function. This representation was found to agree with all known values of the Riemann zeta function once the inherited pole structure from the Bessel zeta function is accounted for. In a later paper the numerical properties of these representations will be investigated as well as an examination of the case of $\nu = -1/2$, which will generate a representation of a Hurwitz zeta function.

## CONFLICTS OF INTEREST

The authors have no conflicts of interest to disclose.

## DATA AVAILABILITY

Data sharing is not applicable to this article as no new data were created or analyzed in this study.

## APPENDIX: RECOVERY OF RIEMANN ZETA FUNCTION

To recover a representation of the Riemann zeta function use Eqs. (2) and (97) with $\nu = 1/2$, then,

$$\zeta(s) = \pi^{s-1} \sin\left(\frac{s\pi}{2}\right)\left(\sum_{n=1}^{\infty}\frac{(-1)^{n+1}\alpha_n s}{2n - s} + \frac{s}{s-1} + \beta_0 + \sum_{n=1}^{\infty}\frac{\beta_n s}{n+s} - \frac{1}{s}\right). \quad (126)$$

This representation can be seen to inherit the pole at $s = 1$, with the correct residue. The Bessel zeta function also has poles at $s = 1 - 2k$, for $k = 1,2,3,...$ so these points are outside the domain for this representation of the Riemann zeta function. For $\nu = 1/2$ the $\alpha_n$ can be expressed in terms of the Bernoulli numbers, combining Eqs. (61) and (82), and recalling that all but $\beta_0$ of the $\beta_n$ are zero, that is,

$$\alpha_n = \frac{(-1)^{n+1}}{n \cdot 2^{2n+1}} a_n = (-1)^{n+1} \frac{2^{2n-1}}{n(2n)!} B_{2n}, \quad (127)$$

$$\beta_0 = -\ln(2), \quad (128)$$

$$\beta_{2n} = -\frac{c_{2n-1}}{2n} = 0, \quad (129)$$

$$\beta_{2n-1} = -\frac{c_{2n-2}}{2n-1} = 0, \tag{130}$$

then,

$$\zeta(s) = \pi^{s-1} \sin\left(\frac{s\pi}{2}\right) \left( \sum_{n=1}^{\infty} \frac{2^{2n-1} B_{2n} s}{n(2n)!(2n-s)} + \frac{s}{s-1} - \ln(2) - \frac{1}{s} \right). \tag{131}$$

This representation has the correct value at the origin. Keeping only the term that survives the limit gives,

$$\zeta(0) = \frac{1}{\pi} \lim_{s \to 0} s \frac{\pi}{2}\left(-\frac{1}{s}\right) = -\frac{1}{2}. \tag{132}$$

It should also be noted that the slope of Eq. (131) agrees with the known value at the origin as well.

$$\zeta'(0) = -\frac{1}{2} \ln(2\pi). \tag{133}$$

This can easily be seen by using Eqs. (2) and (112) or by differentiating Eq. (131) and taking the limit as $s \to 0$.

Now check the value of Eq. (131) at the positive even integers,

$$\zeta(2k) = \lim_{s \to 2k} \pi^{s-1} \sin\left(\frac{s\pi}{2}\right) \left( \sum_{n=1}^{\infty} \frac{2^{2n-1} B_{2n} s}{n(2n)!(2n-s)} + \frac{s}{s-1} - \ln(2) - \frac{1}{s} \right). \tag{134}$$

The only term that will survive the limit is in the sum,

$$\zeta(2k) = \pi^{2k-1} \lim_{s \to 2k} \sin\left(\frac{s\pi}{2}\right) \sum_{n=1}^{\infty} \frac{2^{2n-1} B_{2n} 2k}{n(2n)!(2n-s)}. \tag{135}$$

Now let $s = t + 2k$

$$\zeta(2k) = \lim_{t \to 0} \pi^{2k-1} \sin\left(t\frac{\pi}{2} + k\pi\right) \sum_{n=1}^{\infty} \frac{2^{2n-1} B_{2n} 2k}{n(2n)!(2n-t-2k)}, \tag{136}$$

or,

$$\zeta(2k) = \pi^{2k-1}(-1)^k \lim_{t \to 0} \sin\left(t\frac{\pi}{2}\right) \sum_{n=1}^{\infty} \frac{2^{2n-1} B_{2n} 2k}{n(2n)!(2n-t-2k)}. \tag{137}$$

Keeping the only term in the sum that survives the limit gives,

$$\zeta(2k) = \pi^{2k-1}(-1)^k \lim_{t \to 0} t \frac{\pi}{2} \frac{2^{2k} B_{2k} k}{k(2k)!(-t)}, \tag{138}$$

or,

$$\zeta(2k) = (2\pi)^{2k}(-1)^{k+1} \frac{B_{2k}}{2(2k)!}. \tag{139}$$

This agrees with the known result[17]. Using Eq. (21) and noting that $c_0 = 0$ when $\nu = 1/2$ it can easily be seen that Eq. (130) is zero when $s = -2k$, for $k = 1,2, ...$, which agrees with the known result.

$$\zeta(-2k) = 0. \tag{140}$$

The last remaining identity to verify is,

$$\zeta(1 - 2k) = -\frac{B_{2k}}{2k}, \quad \text{for } k = 1,2,3, .... \tag{141}$$

Unfortunately, the points $s = 1 - 2k$, for $k = 1,2,3, ...$ are outside of the domain for this representation of the Riemann zeta function. There is a subtlety in this case to consider; recall that $s = 1 - 2k$ for $k = 1,2,3, ...$ are the locations of the poles for the Bessel zeta function. Also note that the numerators on the terms generating the poles go to zero when the order of the Bessel function goes to 1/2. To obtain a representation of the Riemann zeta function from the Bessel zeta function the following limit is used,

$$\zeta(s) = \pi^s \lim_{\nu \to \frac{1}{2}} \zeta_\nu(s). \tag{142}$$

To attempt to evaluate this at $s = 1 - 2k$ for $k = 1,2,3, ...$ the following limit would be used,

$$\zeta(1 - 2k) = \pi^{1-2k} \lim_{\substack{\nu \to \frac{1}{2} \\ s \to 1-2k}} \zeta_\nu(s). \tag{143}$$

Explicitly, these two limits are,

$$\zeta(1 - 2k) = \pi^{-2k} \left( \lim_{\substack{\nu \to \frac{1}{2} \\ s \to 1-2k}} \sin\left(\frac{s\pi}{2}\right) \left( \sum_{n=1}^{\infty} \frac{(-1)^{n+1} \alpha_n s}{2n - s} + \frac{s}{s-1} + \beta_0 + \sum_{n=1}^{\infty} \frac{\beta_n s}{n+s} \right. \right.$$
$$\left. \left. - \left(\nu + \frac{1}{2}\right)\frac{1}{s} \right) \right). \tag{144}$$

Pulling out terms where the limit is clear, and no zeros or infinities occur gives,

$$\zeta(1-2k) = \frac{(-1)^k}{\pi^{2k}}\left(1 + \frac{1}{4k^2-2k} - \ln(2) + \sum_{n=1}^{\infty}\frac{2^{2n}B_{2n}}{2n(2n)!}\frac{1-2k}{2n+2k-1}\right.$$
$$\left. + \lim_{\substack{\nu\to\frac{1}{2}\\s\to 1-2k}}\sum_{n=1}^{\infty}\frac{\beta_n s}{n+s}\right). \tag{145}$$

This leaves just one limit to evaluate,

$$\lim_{\substack{\nu\to\frac{1}{2}\\s\to 1-2k}}\sum_{n=1}^{\infty}\frac{\beta_n s}{n+s} = (1-2k)\lim_{\substack{\nu\to\frac{1}{2}\\s\to 1-2k}}\sum_{n=1}^{\infty}\frac{\beta_n}{n+s}. \tag{146}$$

Referring to Eqs. (87)-(89) and noting that the $\beta$'s go to zero if $\nu$ goes to 1/2 and that the denominator goes to zero when $n = 2k-1$, the only term in the series will survive the limit is,

$$\lim_{\substack{\nu\to\frac{1}{2}\\s\to 1-2k}}\sum_{n=1}^{\infty}\frac{\beta_n}{n+s} = \lim_{\substack{\nu\to\frac{1}{2}\\s\to 1-2k}}\frac{\beta_{2k-1}}{2k-1+s}. \tag{147}$$

All pieces of the remaining limit are well defined; however, it is not possible to evaluate the limit as the numerator and denominator each go to zero with respect to different parameters. This is because the original zeta function, Eq. (97), has poles at $s = 1 - 2k$. Hence, it should not be expected that the representation, Eq. (131), be defined at the points $s = 1 - 2k$ for $k = 1,2,3,\ldots$. Fortunately, these points are isolated, and the correct value of the Riemann zeta function is known at the points in question. This gives a final representation for the Riemann zeta function, as obtained from the Bessel zeta function,

$$\zeta(s) = \begin{cases} \pi^{s-1}\sin\left(\frac{s\pi}{2}\right)\left(\sum_{n=1}^{\infty}\frac{2^{2n-1}B_{2n}s}{n(2n)!(2n-s)} + \frac{s}{s-1} - \ln(2) - \frac{1}{s}\right), & s \neq -1, -3, \ldots \\ -\frac{B_{1-s}}{1-s}, & s = -1, -3, \ldots \end{cases} \tag{148}$$

**REFFERENCES**